\documentclass[prl, twocolumn, superscriptaddress]{revtex4-1}
\usepackage[utf8]{inputenc}
\usepackage{amsmath}
\usepackage{amsfonts}
\usepackage{bm}
\usepackage{amssymb}
\usepackage{graphicx}
\usepackage{ dsfont }

\begin{document}

\author{Andrea De Lucia}
\affiliation{Institute of Physics, Johannes Gutenberg University Mainz, Staudingerweg 7, 55128 Mainz, Germany}
\affiliation{Graduate School of Excellence - Materials Science in Mainz, Staudingerweg 9, 55128 Mainz, Germany}

\author{Kai Litzius}
\affiliation{Institute of Physics, Johannes Gutenberg University Mainz, Staudingerweg 7, 55128 Mainz, Germany}
\affiliation{Graduate School of Excellence - Materials Science in Mainz, Staudingerweg 9, 55128 Mainz, Germany}
\affiliation{Max Planck Institute for Intelligent Systems, Heisenbergstrasse 1, 70569 Stuttgart, Germany}

\author{Benjamin Krüger}
\affiliation{Institute of Physics, Johannes Gutenberg University Mainz, Staudingerweg 7, 55128 Mainz, Germany}

\author{Oleg A. Tretiakov}
\affiliation{Institute for Materials Research, Tohoku University, Sendai 980-8577, Japan}
\affiliation{School of Natural Sciences, Far Eastern Federal University, Vladivostok 690950, Russia}

\author{Mathias Kläui}
\affiliation{Institute of Physics, Johannes Gutenberg University Mainz, Staudingerweg 7, 55128 Mainz, Germany}
\affiliation{Graduate School of Excellence - Materials Science in Mainz, Staudingerweg 9, 55128 Mainz, Germany}

\title{Multiscale simulations of topological transformations in magnetic Skyrmions}

\begin{abstract}
Magnetic Skyrmions belong to the most interesting spin structures for the development of future information technology as they have been predicted to be topologically protected. To quantify their stability, we use an innovative multiscale approach to simulating spin dynamics based on the Landau-Lifshitz-Gilbert equation. The multiscale approach overcomes the micromagnetic limitations that have hindered realistic studies using conventional techniques. We first demonstrate how the stability of a Skyrmion is influenced by the refinement of the computational mesh and reveal that conventionally employed traditional micromagnetic simulations are inadequate for this task. Furthermore, we determine the  stability quantitatively using our multiscale approach. As a key operation for devices, the process of annihilating a Skyrmion by exciting it with a spin polarized current pulse is analyzed, showing that Skyrmions can be reliably deleted by designing the pulse shape.
\end{abstract}

\maketitle

%\section{Introduction}
Magnetic Skyrmions \cite{Skyrme} are topological spin structures that arise in the spin pattern of ferromagnetic systems with broken inversion symmetry, such as chiral crystals \cite{chimag1, chimag2} or thin magnetic films with different top and bottom interfaces \cite{thinfilm1, thinfilm2}. Skyrmion lattices \cite{SkLat1, SkLat2, SkLat3, woo} constitute the ground state for some systems, while isolated Skyrmions can appear as metastable states of some magnetic nanostructures \cite{nanodisk}. Isolated Skyrmions have been recently considered \cite{isSk1, isSk2, finocchio, kainew} as the building blocks for ultradense magnetic storage devices \cite{Kra-Wie}.

Skyrmions carry a topological charge $Q= \pm 1$ defined as \cite{WDSMS}:
\begin{equation} \label{eq:skynum}
Q = \frac{1}{4 \pi}\int_{A}  \, \mathrm{\textbf{m}} \cdot \left(\frac{\partial \mathrm{\textbf{m}}}{\partial x} \times \frac{\partial \mathrm{\textbf{m}}}{\partial y} \right)dx dy,
\end{equation}
where $A$ is the area of the system and $\mathrm{\textbf{m}}$ the unit magnetization vector.
Since transitions that change $Q$ are forbidden \cite{WDSMS} in a continuum description of $\mathrm{\textbf{m}}$, such structures are topologically protected. Nevertheless, in a real system composed of discrete magnetic moments localized on the atomic lattice sites, no strict topological protection exists \cite{Gar-Chud}. Thus it is necessary to overcome a finite energy barrier to induce transformations that change $Q$, such as the annihilation of a Bloch line (BL) \cite{WDSMS, slonc, sovpaper, BL1, BL2, Moutafis, patent, ortuno}. 

The stability against external fields is indeed a key feature of Skyrmions, making them a good candidate as information carriers in next generation storage devices \cite{skytech1, skytech2, skytech3}. The fundamental prerequisites for applications are ascertaining the stability of Skyrmions, as well as reliably annihilating them. However, the computational treatment of processes involving annihilating Skyrmions is very delicate. In analytical micromagnetic theory, singularities in the exchange field tend to arise during topological transformations, making numerical simulations very susceptible to the mesh being used \cite{thiav_mesh} and therefore often inaccurate. The necessity for a computational model, capable of performing quantitatively accurate simulations is therefore obvious and a key step. While more accurate atomistic simulations would overcome this problem, the computational power required to run such simulations for a sample of realistic experimental size makes this possibility infeasible.

In this work, the annihilation of isolated Skyrmions is studied by simulating the Landau-Lifshitz-Gilbert (LLG) equation with a multiscale approach \cite{io}. Within this approach the core of the Skyrmion is simulated  atomistically, while the remaining part of a nanodisk hosting the Skyrmion is simulated using micromagnetics. This technique was designed to ensure computational accuracy combined with feasible computational times.

First, the effects of lattice on the stability are studied, showing how the mesh density influences the annihilation of Skyrmions. Then, BLs are excited along the domain wall separating the two out-of-plane magnetized domains in the Skyrmion. For this purpose we employ current pulses that generate spin-orbit torques \cite{fund_spinorb, Hayashi, Olegnew}, showing how the shape of these pulses influences the Skyrmion and induces changes to the topology. Finally, we analyze how we can reliably annihilate the Skyrmions by tailoring the pulse shape, which thus presents a quick and robust way to delete selected Skyrmions.

%\section{method and results}
The simulations were performed in a ferromagnetic disk with the radius of 53 nm and thickness of 3 nm, using the saturation magnetization $M_{s} = 10^{6}$ A/m, out-of-plane anisotropy constant $K_{z} = 1.3 \times 10^{6}$ J/m$^3$, and the exchange constant $A = 1.1 \times 10^{-11}$ J/m. These parameters are comparable to those of CoFeB \cite{CoFeB} in multilayer stacks that are widely used in thin film nanostructures that exhibit Skyrmions \cite{woo}.  The central part of the system was simulated atomistically (fine scale region), while the remaining part was simulated using the micromagnetic model (coarse scale region), following the approach described in Ref.~\cite{io}. The size of the fine scale region was chosen to fit the entire Skyrmion at rest, but without sacrificing too much computational time. It should be stressed that larger Skyrmions can still be simulated accurately as far as no discontinuities occur in the coarse scale region. First, magnetic Néel Skyrmion states were relaxed for different values of the Dzyaloshinskii-Moriya interaction (DMI), then simulations with a constant uniform magnetic field, applied in the direction opposite to the magnetization inside the Skyrmion were performed. All the micromagnetic parameters were kept fixed, whereas the atomistic ones were changed. In particular, the distance $a$ between two neighboring nodes of the mesh was changed, in order to increase the density of spins. While $a$ can be interpreted as the lattice constant of the material, it is treated in this case just as a computational parameter. As a result,  the magnetic moment of the spins $\mu$ and the exchange constant $J$ were rescaled according to $\mu = a^3 M_s$ and $J = aA$. This effectively simulates materials that are consistent with the same micromagnetic parameters, which are used for the coarse scale region.

\begin{figure}[hbtp]
\centering
\includegraphics[width=0.8\columnwidth]{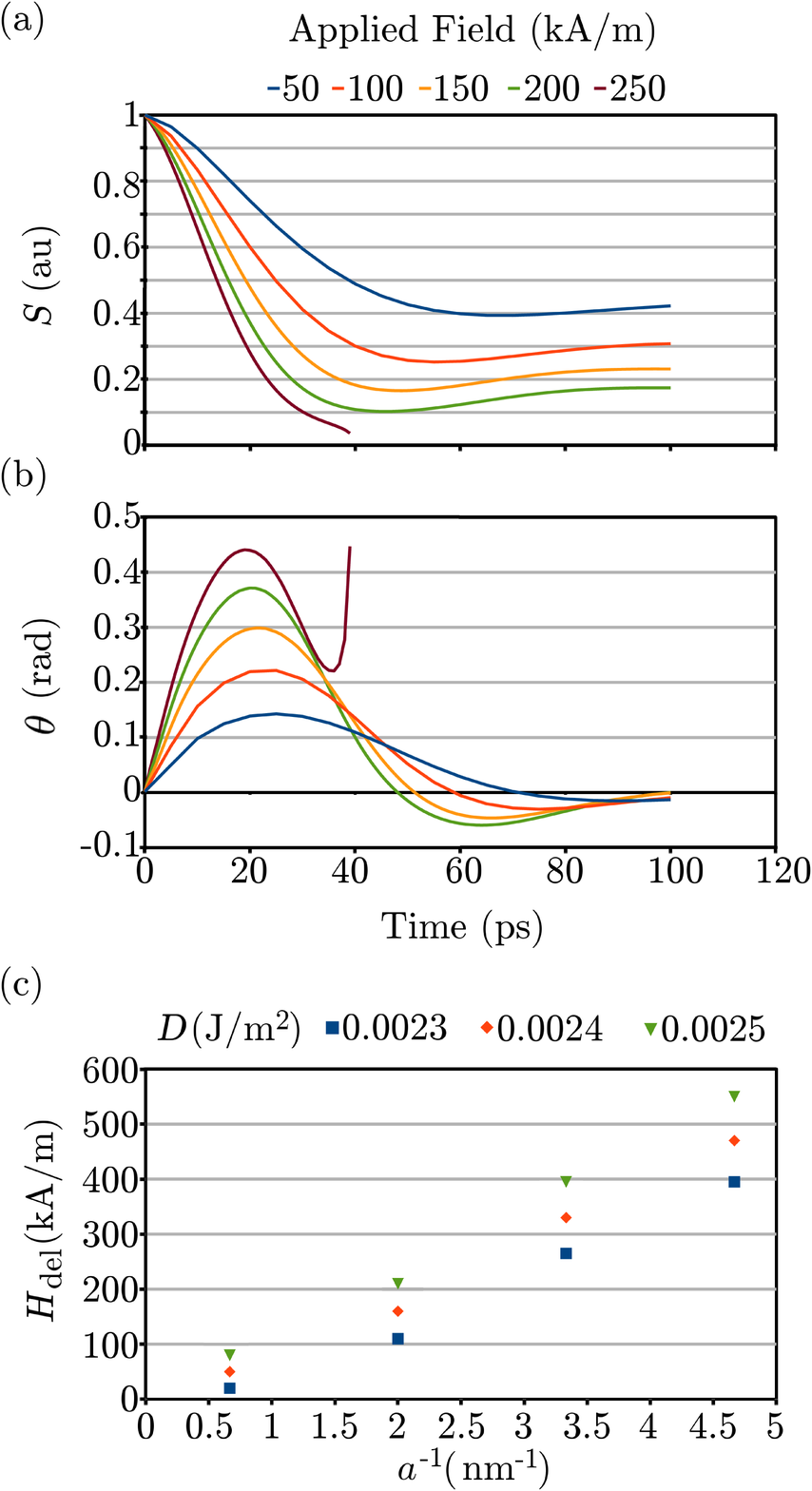}
\caption{Dynamics of a Skyrmion for different values of a constant out-of-plane field. The system shows an oscillatory behavior, where both (a) its size, expressed in terms of the Skyrmion magnetic moment $S$, and (b) the angle $\mathrm{\theta}$ between the in-plane magnetization components of the domain wall and the radial direction, reach a certain nonzero value before relaxing back to the equilibrium. The data corresponding to 250 kA/m shows the Skyrmion annihilation. (c) The minimum magnetic field $H_{del}$ necessary to adiabatically annihilate a Skyrmion for different values of the DMI constant $D$ and linear spin density $a^{-1}$. Since finer meshes lead to higher values of $H_{del}$, the case for an indestructible Skyrmion in a continuous model can be considered as a limiting case. $H_{del}$ can be shown to linearly increase as a function of the spin density. Data points corresponding to the lowest value of $a^{-1}$ were simulated in purely micromagnetic simulations.
\label{fig:rad}}
\end{figure}

\begin{figure}[hbtp]
\centering
\includegraphics[width=0.9\columnwidth]{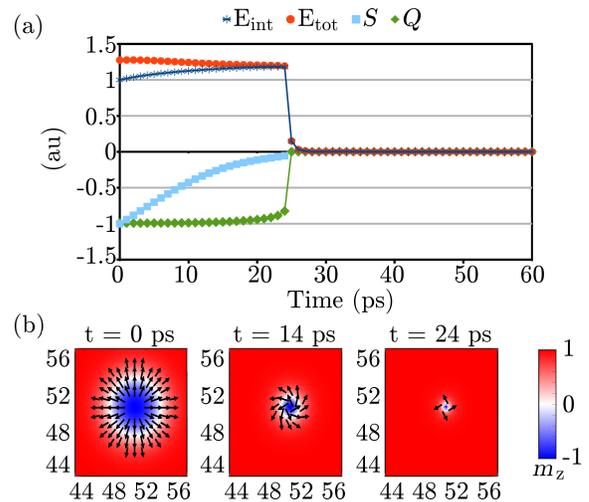}
\caption{(a) Path to annihilation of a Skyrmion in 300 $\mathrm{kA/m}$ external magnetic field. The internal energy $E_{int}$ of the system and the total energy (internal plus Zeeman energies) are compared as functions of time. A potential barrier exists for the internal energy which has to be overcome by the application of an external field. All the quantities are presented in arbitrary units. (b) Dynamic snapshots at various stages of the annihilation process. The initial configuration of a Néel Skyrmion is perturbed when the structure is shrinking. The scale on the axes is expressed in units of the micromagnetic computational cell (3 nm). \label{fig:ecomp}}
\end{figure}

In Fig.~\ref{fig:rad}(a) we show that an application of an external out-of-plane magnetic field leads to the Skyrmion shrinking until it reaches its new equilibrium size. This behavior is reproduced for magnetic fields up to critical value $H_{del}$. For fields larger $H_{del}$, the Skyrmion shrinks until it completely annihilates. The analysis of the Skyrmion dynamics in nonzero out-of-plane fields shows that the spins magnetized in plane, corresponding to the center of the Skyrmion's circular domain wall, tilt clockwise while the Skyrmion shrinks [see Figs.~\ref{fig:rad}(b) and \ref{fig:ecomp}(b)]. When the shrinking stops, i.e. the Skyrmion reaches a new equilibrium size, the magnetization in the domain wall aligns along the radial direction again, recovering the Néel Skyrmion character. 

One of the measures of the Skyrmion size can be the total magnetic moment $S$ inside the Skyrmion's domain wall. It is proportional to $\sum_{i} \left(m_{z,i}-1\right) $, where $m_{z,i}$ is the out-of-plane component of the normalized magnetization at the lattice site $i$, and the sum runs over all cites in the fine scale region, which always completely includes the skyrmion's domain wall. For fields below $H_{del}$, we observe that $S$ reaches a minimum, depending on Gilbert damping $\alpha$, before relaxing back to a slightly larger value [Fig.~\ref{fig:rad}(a)]. While the Skyrmion increases in size the magnetization in the domain wall tilts counterclockwise [Fig.~\ref{fig:rad}(b)]. The aim here is to demonstrate how the simulation results can be influenced by the refinement of the mesh rather than testing the stability of the Skyrmion for different material parameters, as previously investigated e.g. in Ref.~\cite{Gar-Chud}.

We find that with decreasing $a$, i.e. increasing the density of magnetic moments, leads to an increase of $H_{del}$ [see Fig.~\ref{fig:rad}(c)]. This is in agreement with Ref.~\cite{Siemens} and shows how the minimum size which a Skyrmion can reach before the annihilation strongly depends on the lattice constant. Furthermore, these results agree with the asymptotic behavior of indestructible Skyrmions in a continuous model. Topological protection can thus be considered a limiting case of the energy barrier \cite{Gar-Chud} separating the metastable Skyrmion state from the ferromagnetic ground state. 

The energy barrier is shown in Fig.~\ref{fig:ecomp}(a), where the internal energy $E_{int}$ of a Skyrmion shrinking under the influence of a constant magnetic field is plotted as a function of time. It can be noticed that $E_{int}$, consisting of the exchange, anisotropy, dipolar energy, and DMI contribution, increases until the annihilation occurs. The energy barrier is overcome by the application of the Zeeman energy. The Skyrmion moment $S$ is also shown, to stress that once the Skyrmion reaches its minimum size, the topological barrier is overcome, and the system relaxes in the more stable uniform ferromagnetic state. It can be further noticed that the Skyrmion charge $Q$ instantly switches to zero when the barrier is overcome.
%It should be stressed, that the purpose of Fig.~\ref{fig:rad}(c) is to show how the stability of the Skyrmion is influenced by the refinement of the mesh, rather than comparing different materials.
A purely micromagnetic simulation with a $\mathrm{1.5}$ nm cell size ($a^{-1} \simeq 0.667$ nm$^{-1}$) was included for comparison. It cannot be understated that a multiscale approach is able to simulate the singularities atomistically using realistic material parameters, and the uniformly magnetized external region in the micromagnetic model, allows one to predict the dynamics of a similar system with better quantitative accuracy than obtainable using only the micromagnetic model.

Unlike the atomistic model, where the correct lattice constant must be used in order to obtain realistic results, the micromagnetic model becomes more and more accurate by refining the mesh. Ideally, in the infinitely fine mesh limit the analytical theory is recovered. Nevertheless even in the analytical theory, predictions made by the micromagnetic model can be in disagreement with the experimental evidence. These intrinsic limitations are derived from the micromagnetic model neglecting of the length scales comparable to the lattice constant. The magnetization vector itself, which is the fundamental quantity that the model investigates, is proportional to the local average of the atomic magnetic moments. According to the definition \cite{Menc-Silv} $$\mathbf{M} = \lim_{\tau \to 0}\frac{1}{\tau} \sum_{i}^{N} \mathbf{\mu}_i = \lim_{\tau \to 0}\frac{N}{\tau}  \left\langle \mathbf{\mu} \right\rangle $$ where $\tau$ indicates a volume element containing $N$ magnetic moments $\mathbf{\mu}$. The limit $\tau \to 0$ should be considered to be restricted to the volume of elements which are small compared to the full magnetic system but large enough to contain a statistically significant number of magnetic moments. One basic example is the excitation of spin waves with a wavelength smaller than the lattice constant, a phenomenon that does indeed arise in a continuum model despite being forbidden in experiments and in realistic atomistic simulations. This thus shows that only a multiscale simulation can reproduce the dynamics realistically.

As the annihilation of a Skyrmion includes the annihilation of a BL, this phenomenon cannot be properly simulated in the micromagnetic framework. Because of this change in topology of the spin structure during the process \cite{sovpaper}, charge $Q$ of the structure changes from $\mathrm{\pm 1}$ to 0, thus lifting the topological protection. Extending the results of Ref.~\cite{Moutafis}, where a BL is formed and annihilated in a Bloch Skyrmion via application of a field gradient, we study a similar singularity generated by a spin-polarized current pulse applied along the $x$-direction. In general, using spin currents is more advantageous than using fields to manipulate magnetization due to more favorable scaling. The influence of the spin-orbit torques on Skyrmions \cite{Olegnew}, in particular yields many promising possibilities towards the implementation of Skyrmions as information bits. The LLG equation implemented to include the effect of a spin-polarized current (generated for instance via the inverse spin galvanic effect or the spin-Hall effect) \cite{Hayashi} reads: 
\begin{equation} \label{eq:shtLLG}
\begin{split}
\frac{d \mathbf{m}}{d t} = &-\gamma^{\prime} \left[ \mathbf{m} \times  \mathbf{H}_{eff} + \alpha \left( \mathbf{m} \times \left(\mathbf{m} \times \mathbf{H}_{eff} \right)\right) \right] \\ &- \gamma^{\prime}a_J \left[ \left(\xi - \alpha  \right) \left( \mathbf{m} \times \mathbf{p}\right) + \left(1 + \alpha \xi \right) \left(\mathbf{m} \times  \left(\mathbf{m} \times \mathbf{p}\right)\right) \right],
\end{split}
\end{equation}
where $\gamma^{\prime} = {\gamma}/\left(1 + \alpha^2\right)$ with $\gamma$ being the gyromagnetic ratio, $\mathbf{H}_{eff}$ is the effective field, $\mathbf{p}$ the average polarization of the current generated by the spin-Hall effect, $a_J$ the damping-like term constant \cite{dampinglike1, dampinglike2, fieldlike}, and $\xi$ the ratio between damping-like and field-like torques. We apply the current density pulses of Gaussian shape, $J (t) = J_0 \exp [-t^2/(2 \sigma^2)]$.

We find that the annihilation of BLs can be excited in a Néel Skyrmion for some combinations of $ J_0 $ and $\mathrm{\sigma}$. The material parameters employed for this simulation were the same as the stability simulations with the damping constant $\alpha = 0.1$. Furthermore the Hall angle $\alpha_{H} = 0.1$ and the constant $\xi = 0.5$ were used.

\begin{figure}[hbtp]
\centering
\includegraphics[width=0.9\columnwidth]{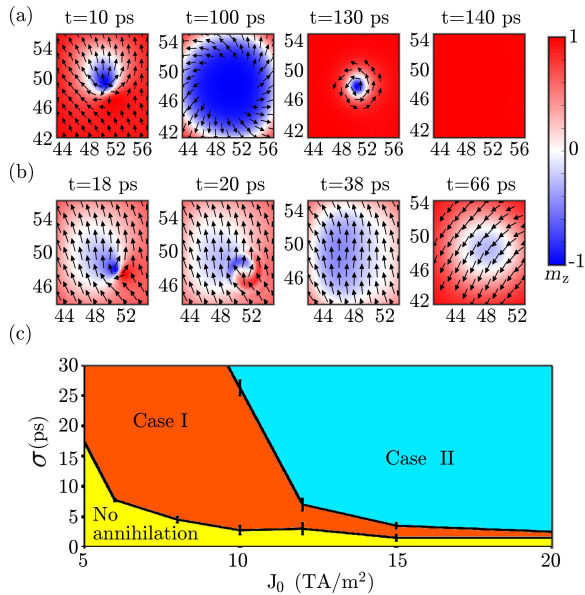}
\caption{Spin structure of the Skyrmion during the annihilation process in cases I and II. The color code for the arrows shows the out-of-plane component, from red, to white to blue. The initial state corresponds to a relaxed Skyrmion centered on the cell with coordinates (50, 50). (a) Case I: A BL is formed in the domain wall of the skyrmion within a vortex-antivortex pair. The spins in the domain wall turn clockwise starting from the position of the pair, meanwhile, the Skyrmion increases in size and reaches a maximum then starts shrinking in size. As the Skyrmion shrinks below the minimum size, it is finally annihilated. The system relaxes into the ferromagnetic ground state.
(b) Case II: The vortex-antivortex pair annihilates, the Skyrmion number immediately turns to zero, the system quickly relaxes back to the ferromagnetic state.
(c) Different regimes depending on the peak height $J_0$ and half-width $\mathrm{\sigma}$ of the Gaussian pulse. The error bars were evaluated by performing simulations with different values of $\sigma$. \label{fig:6pic}}
\end{figure}

The results show that it is indeed possible to form a BL, as a vortex-antivortex couple on the domain wall of the skyrmion, using spin-orbit torques. In the process the domain wall deforms, increases in width on one side of the Skyrmion and decreases on the opposite side. The duration of the pulse plays a fundamental role, since a pulse that is too short would not deform the domain wall enough, while a pulse too long would act adiabatically on the whole Skyrmion and push it beyond the edge of the magnetic system. Intermediate values result in the formation of the BL that can either annihilate or relax. While the annihilation is a topological transformation and leads to the annihilation of the Skyrmion, the relaxation of the BL results in its rapid expansion. We can explain the rapid expansion of the Skyrmion as a consequence of the large exchange energy density of the BL being dissipated as a coherent excitation of the Skyrmion.
It is possible to distinguish three different regimes, see Fig.~\ref{fig:6pic}.
In the nonannihilating regime the relaxation of the BL is accompanied by Skyrmion's size oscillations, which do not lead to collapse. As was noted earlier in the paper, Skyrmions collapse once their size becomes too small to stabilize them in the antiparallelly aligned surrounding magnetization. This occurs in the annihilation regime of case I  [Fig.~\ref{fig:6pic}(a)], which (while not qualitatively different from the nonannihilating regime) results in stronger size oscillations that annihilate the Skyrmion due to overshooting in the shrinking phase.
The annihilation regime of case II, see Fig.~\ref{fig:6pic}(b), is indeed qualitatively different since the vortex-antivortex pair with opposite polarities forms and subsequently annihilates \cite{hertel, oleg}, leading to the immediate annihilation of the Skyrmion. This regime could thus be exploited for practical applications since it allows to lift the topological protection of Skyrmions in a quick and reliable manner.

%\section{conclusions}
In conclusion, we have determined the Skyrmion stability using a multiscale approach that allows for a more realistic description of Skyrmion annihilation compared to conventionally used micromagnetics. We have demonstrated that the stability of Skyrmions is strongly influenced by computational parameters, such as the mesh size. Using the multiscale approach overcomes this problem, and allows one to obtain the realistic Skyrmion stability parameters. The cell size here is fixed by the appropriate lattice constant of the simulated material, and the computational efforts are far lower than those of a purely atomistic simulation. Furthermore, this approach reproduces the dynamics including the spin spectrum realistically, which even allows in the future to include thermal effects. We employ this multiscale approach to study topological transformations by applying spin-orbit torques due to spin-polarized current pulses. This is shown to be a very fast and efficient method to delete isolated Skyrmions as required for applications. We ascertain the combinations of pulse parameters, which robustly annihilate the Skyrmion. This may open up a path to delete Skyrmions reliably as required for future spintronic memory.

%\section{Acknowledgements}
A.\,D.\,L. and K.\,L. are recipients of a scholarship through the Excellence Initiative by the Graduate School Materials Science in Mainz (GSC 266), B.\,K. is the recipient of the Carl Zeiss Postdoc Scholarship – Multiskalensimulationen für energiesparende Magnetisierungsmanipulation. The authors acknowledge the support of SpinNet (DAAD Spintronics network, project number 56268455) and the DFG (SFB TRR 173 SPIN+X). O.\,A.\,T. acknowledges support by the Grants-in-Aid for Scientific Research (Grants No.~25247056 and No.~15H01009) from MEXT, Japan.

\end{document}